\documentclass[aip,apl,preprint,reprint,unsortedaddress,amssymb]{revtex4-1}

\usepackage{graphicx}
\usepackage{textcomp}

\usepackage{savesym}
\usepackage{amsmath}
\savesymbol{iint}
\usepackage{txfonts}
\restoresymbol{TXF}{iint}

\begin{document}
 \title{Deciphering mechanisms of enhanced-retarded oxygen diffusion in doped Si}
 \author{Dilyara Timerkaeva}
 \affiliation{Laboratoire de Simulation Atomistique (L\_Sim), SP2M, INAC, CEA-UJF, 38054 Grenoble Cedex 9, France}
 \affiliation{Kazan Federal University, Institute of Physics, 420018, 18, Kremlevskaya st., Kazan, Russia}
 \author{Damien Caliste}
 \author{Pascal Pochet}
 \affiliation{Laboratoire de Simulation Atomistique (L\_Sim), SP2M, INAC, CEA-UJF, 38054 Grenoble Cedex 9, France}
 
 \begin{abstract}
 In this letter, we study enhanced-retarded diffusion of oxygen in doped silicon by means of first principle calculations.
We found that the migration of oxygen dimers can not be significantly affected  by strain, doping type or rate.
We attribute the enhanced oxygen diffusion in p-doped silicon to reduced monomer migration energy, and the retarded oxygen diffusion in Sb-doped to oxygen trapping close to a dopant site.
The two proposed kinetic and thermodynamic mechanisms can appear at the same time and  might lead to contradictory experimental results.
Such mechanisms can be involved in the light induced degradation phenomenon in solar grade silicon.

 \end{abstract}
 
 \maketitle
 
 Oxygen is an important impurity in crystalline silicon. 
 It does not affect the crystallographic structure of Si since it is an interstitial type impurity.
 Among the advantageous effects of oxygen impurities, we can emphasize the improved mechanical properties of the silicon crystal. \cite{Newman2000}
 The internal gettering \cite{Tice1976} process is another beneficial effect occurring due to the oxygen precipitates.
 However, the presence of oxygen can lead to detrimental effects. In particular, it can affect the properties of photovoltaic and microelectronic devices.
 For example, it has been reported that B-doped silicon forms boron-oxygen complexes which lead to the Light Induced Degradation (LID) effect.\cite{Fischer1973, Schmidt2004, Voronkov2010} 
 Though many experimental and theoretical groups have studied the composition of these defects, their exact arrangement is still debated.\cite{Adey2003, Voronkov2010, Schmidt2004} 
 Nevertheless the role of oxygen diffusion in the B-O complexes formation process is undeniable and it has been suggested\cite{Adey2004} that oxygen diffusion in doubly charged state may occur under illuminated conditions
 
 The oxygen interstitial monomer is known to be the primary diffusion particle at temperatures higher than 700~\textcelsius ~with an activation energy $E_{act}=2.53$ eV  as reviewed by Newman.\cite{Newman2000} 
 An enhanced diffusion is observed at lower temperatures (1.5 eV) which is attributed to the diffusion of oxygen dimer.\cite{Coutinho2000, Lee2002,Giannattasio2005} 
 Although some alternative species have been proposed as diffusing particles (e.g. O-I~\cite{Ourmazd1984}, O-V~\cite{Helmreich1977}, molecular oxygen~\cite{Gosele1982, Gosele1989, Cui2008}, interstitial trimers O$_{3i}$ and longer interstitial chains~\cite{Lee2002}), it was shown that they barely participate in the oxygen diffusion process.\cite{Takeno1998, Cui2008}
 
 Doping could also affect the oxygen diffusion. 
 Wada\cite{Wada1984} has reported the early evidence of oxygen diffusion enhancement and retardation in highly doped silicon through measuring the rate of the thermal donors (TDs) formation.
 TD is an electrically active defect which contains oxygen atoms. 
 It has been shown that TDs formation rate is enhanced in heavily p-doped silicon and reduced in heavily n-doped silicon.~\cite{ Wada1984} 
 Recent studies on oxygen diffusion at low temperature range by means of Dislocation Locking Technique confirm the enhanced oxygen diffusion in heavily p-doped silicon, and reduction in heavily n-doped silicon.\cite{Takeno2000, Murphy2006JAP, Murphy2006MSE, Zeng2011}
 However, the nature of such an enhancement remains unclear. 
 Zeng et al.\cite{Zeng2011} have inferred two  possible causes: \textit{i)} strain, since high doping rates modify the cell parameter of the crystal; \textit{ii)} Fermi energy level, since high doping rates suggests the intrinsic semiconductor will transform into an extrinsic one.
  
\begin{figure}
  \includegraphics[width=1.0\linewidth]{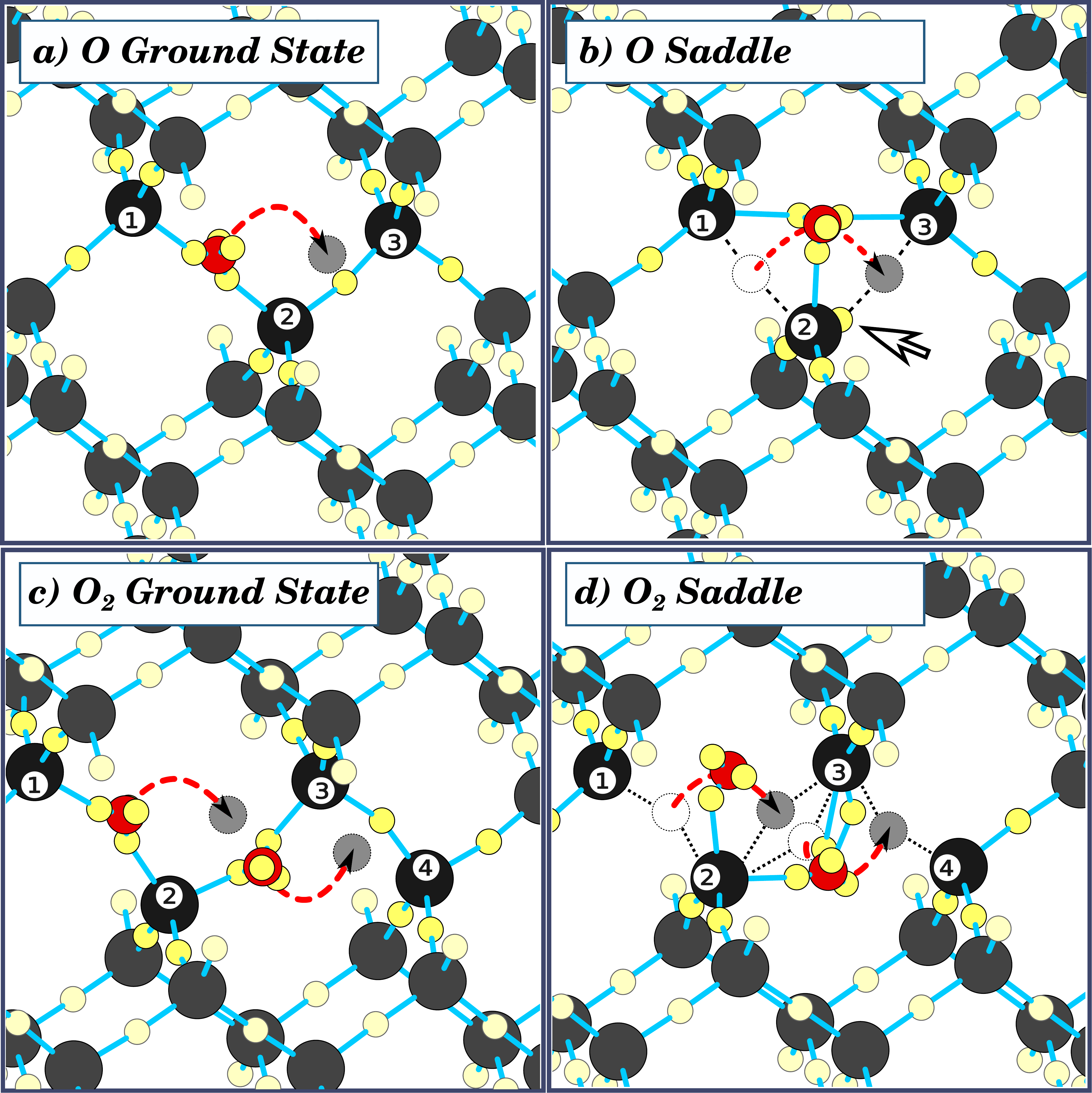}
  \caption{(Color online) O and O$_2$ diffusion schemes. Big black and small red filled balls represent Si and O atoms, respectively. Empty yellow balls represent the positions of the Wannier Centers (see text). The path(s) used by O atom(s) is(are) schematically represented by a dashed arrow pointing to the final position(s) (gray shadows).}
  \label{fig:o-o2}
\end{figure}

 In the present work we aimed to clarify these mechanisms by which doping affects oxygen diffusion in the low temperature regime. 
 The diffusion of O$_i$ at the atomistic level proceeds through the unit jumps along any of $\langle 110 \rangle$ crystallographic direction as illustrated  in Figure \ref{fig:o-o2} \textit{a-b}: oxygen starts to move from the \textit{bond-centered} configuration, passes through the saddle \textit{Y-lid} configuration, and finishes on the neighboring  \textit{bond-centered} configuration.\cite{Newman2000} 
 The diffusion of O$_{2i}$ proceeds in a similar manner \cite{Coutinho2000} (Figure \ref{fig:o-o2} \textit{c-d}): it starts with the \textit{staggered} configuration, passes through the saddle \textit{square} configuration, and finishes with a neighboring \textit{staggered} configuration. 

 The corresponding migration energies (E$_m$) are considered as the energy difference between the saddle (SAD) and the ground state (GS) configurations. These configurations were obtained using Density Functional Theory (DFT) pseudopotential calculations in the Generalized Gradient Approximation. 
 We utilized the wavelet based BigDFT code \cite{Genovese2008} as it is very efficient at dealing with complex and inhomogeneous systems. 
 The calculations were carried out using the Perdew–Burke–Ernzerhof functional within Periodic Boundary Conditions.
 The wavelet basis-set accuracy was given by a grid spacing of 0.44 Bohr. This criterion provides energy values with an accuracy of 20 meV.
   
 Fast inertial relaxation engine\cite{Bitzek2006} (FIRE) and direct inversion in interactive subspace (DIIS)\cite{Pulay1980} force-based optimizers was used to converge to ground state and saddle point configurations, respectively\cite{Machado-Charry2011}. The criterion on forces for geometry optimization was chosen to be smaller than $3\times10^{-4}$ Ha Bohr$^{-1}$.
 
 A first supercell containing 192 Si atoms and with a $1\times1\times3$ $k$-point sampling was employed to simulate the strain effect. 
 X and Y axes of the supercell are collinear to $\langle 110 \rangle $ and $\langle 1\overline{1}0 \rangle$ crystallographic directions, which allowed the application of uniaxial and biaxial strain along the migration path of O$_i$ and O$_{2i}$. Strains ranging from $-1.0$\% to $+1.0$\% have been applied. 
 For both O$_i$ and O$_{2i}$, we observed that the maximal resulting change in E$_m$ is as low as 0.17~eV in this wide range of strain. 
 Moreover, experimental strain values are lower than $0.001$\%\cite{Zeng2011} and therefore create a negligible effect on the migration energies.
 Thus, we can reject the  first hypothesis of Zeng \textit{et al.}\cite{Zeng2011} that attributes the enhancement or retardation of O diffusivity in heavily doped silicon to elastic strain.
 
 To study the second hypothesis of Zeng \textit{et al.},\cite{Zeng2011} we used a larger supercell containing 512 Si atoms and with only the $\Gamma$-point. 
 With such supercells  we aimed to reduce the spurious interactions between the dopant atom, diffusing oxygen, and their periodic images. 
 The relative positions of the defects and the dopant atoms were chosen in order to minimize the elastic interactions within the supercell. 
 The migration energies of an oxygen monomer and dimer were studied in the presence of p-type (B, Al and Ga), n-type (P, As and Sb), and isovalent (Ge) dopant atoms.
 The analysis of the electronic structure of the systems was made in terms of The Maximally Localized Generalized Wannier Functions Code\cite{Mostofi2008} interfaced with BigDFT, which provides accurate pictures of charge distribution and bonding \cite{Boulanger2013}.
 
 We start with the analysis of the O$_i$ migration path in terms of the valence electronic configuration around the oxygen defect. 
 The centers of the Wannier functions (WC)  are depicted in Fig. \ref{fig:o-o2}~\textit{a-b}.
 In the GS configuration, the oxygen atom is bonded to two neighboring Si atoms (Si$^1$ and Si$^2$).
 The remaining valence electrons form two lone pairs on the oxygen.  
 In the SAD configuration, the oxygen is bonded to three Si atoms (Si$^1$, Si$^2$, and Si$^3$). Two of the remaining valence electrons form a lone pair on the oxygen while the last electron is localized in a lone pair near the Si$^2$ atom (arrow in Fig.~\ref{fig:o-o2}~\textit{b}). 
 Next we analyze the O$_{2i}$ migration path as illustrated in Figure \ref{fig:o-o2}~\textit{c-d}. 
 Both oxygen atoms diffuse in the same $\langle110\rangle$ plane and are bonded to their neighboring Si atoms.
 In the GS configuration each has a couple of lone pairs. 
 The electronic distribution of saddle configuration differs from O$_i$ case. 
 It is more complex with three lone pairs on the first oxygen and two lone pairs on the second oxygen. 
 This asymmetry is consistent with the \textit{Mexican hat} shaped diffusion barrier for O$_{2i}$.\cite{Adey2004} 
 
 Having these mechanisms in mind, we now consider the effect of n- and p-type dopants on the migration of O$_{2i}$ by substituting one Si atom by a doping species.
 The values are summarized in Table~\ref{tab:o_eact}. 
 Donor, isovalent and acceptor doping slightly raise E$_m$ by 0.16-0.18~eV, 0.05~eV, and 0.04-0.05~eV, respectively. 
 Such small variations in E$_m$(O$_{2i}$) cannot explain either the experimentally observed enhancement or retardation.
 At the same time, the dopants slightly affect the binding energy of oxygen dimers keeping it between 0.10 and 0.12~eV.
 Only Sb doping led to a significant reduction of the binding energy that could reduce the concentration of O$_{2i}$.
 For the other dopants, it is relevant to study the doping effect on the monomer itself as it is the precursor to the formation of O$_{2i}$. 
 \begin{table*}
    \begin{tabular}{cccccc}
        \hline \hline
        dopant element & $E_{bind}(O_{2i})$ & $E_{m}(O_{2i}), eV$ & $\delta E_m(O_{2i}), eV$&  $E_{m}(O_i), eV$ & $\delta E_m(O_i), eV$ \\
         \hline \hline
         no dopant &  0.10 & 1.58 & ref & 2.41 & ref.\\
         \hline 
         B  & 0.12 & 1.63 & 0.05 & 2.06 & -0.35 \\
         BB &  \textemdash  & \textemdash & \textemdash & 1.98 & -0.43 \\
         Al & 0.12 & 1.63 & 0.05 & 2.04 & -0.37 \\
         Ga & 0,11 & 1.62 & 0.04 & 2.05 & -0.36 \\
         \hline
         Ge & 0.11 & 1.63 & 0.05 & 2.39 & -0.02 \\
         \hline
         P  & 0.11 & 1.75 & 0.17 & 2.26 & -0.15 \\
         As & 0.10 & 1.76 & 0.18 & 2.29 & -0.12 \\
         AsAs & \textemdash & \textemdash & \textemdash & 2.30 & -0.11 \\
         Sb & -0.39 & 1.74 & 0.16 & 2.24 & -0.17 \\
        \hline \hline
    \end{tabular}
  \caption{First data column represents the binding energies of the $O_{2i}$ aggregate in doped or undoped supercells. The other columns are for the migration energies of the $O_{2i}$ and $O_{i}$ species and relative variations in the presence of dopants. p-doping represents the heavier variations of O monomer migration energy while all other values are less impacted.}
  \label{tab:o_eact}
 \end{table*}
 The effect of dopants on O$_i$ diffusion is summarized in Table \ref{tab:o_eact}. 
 We observe a more significant impact for p-type dopants than for n-type dopants. The reduction of E$_m$ is by 0.35-0.37~eV in the former case, whereas we observe a reduction of E$_m$ by 0.11-0.16~eV in the latter case. 

 \begin{figure*}[t]
  \includegraphics[width=1.0\linewidth]{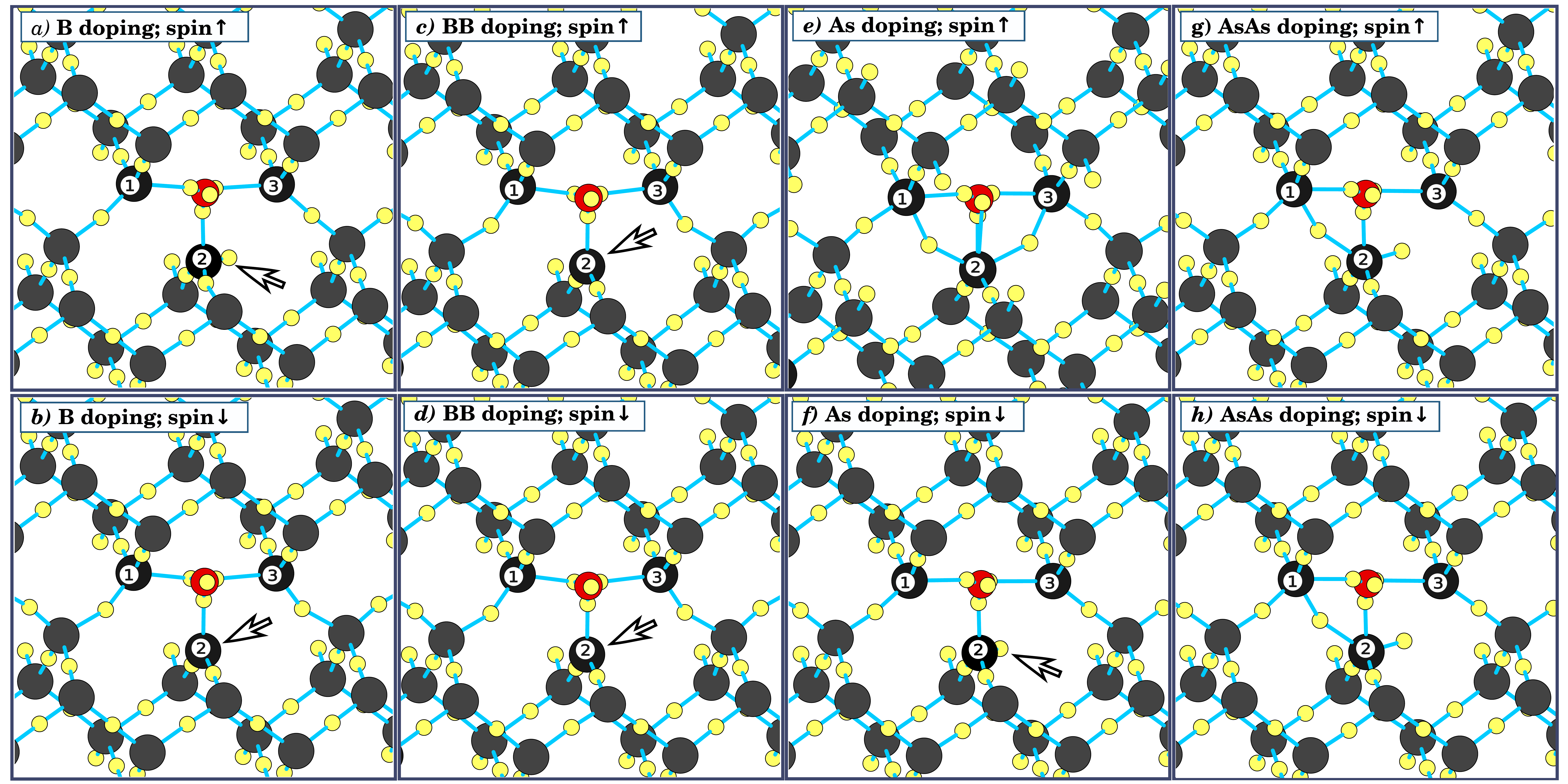}
  \caption{(Color online) Saddle configurations of O$_i$ monomer with additional dopant(s) in the supercell a$)$ B b$)$ B + B c$)$ As d$)$ As + As. Atoms and Wannier centers use the same schematic conventions as Figure~1. The upper line of configurations represents spin up while the line below is used for the spin down channel. We can observe that B-doping suppresses lone-pairs around Si$_2$, while As-doping creates additional bonding between Si$_1$, Si$_2$ and Si$_3$.}
 \label{fig:dopant_sad}
 \end{figure*}

 We now use the Wannier analysis to get physical insight for this migration energy reduction.
 For the GS configuration, the electronic distribution around oxygen is similar to the intrinsic case for all tested dopants as shown in Figure~\ref{fig:o-o2}~\textit{a}. 
 Thus the effect should lie in the SAD configuration.
 We took As and B as prototypes for both types of doping. Spin polarized calculation have been used and both spin up (SU) and spin down (SD) are depicted.
 The rearrangement of the electronic structure of oxygen in the \textit{Y-lid} configuration is reported in Fig.~\ref{fig:dopant_sad}.
 In the B-doped case, the relative arrangement of WCs with SU is very similar to the intrinsic case \textendash~a lone pair is near the Si$^2$ atom, while it is not the case for the SD electrons (Fig.~\ref{fig:dopant_sad}~\textit{a-b}). 
 Indeed, the corresponding SD center of Si$^2$ is now empty as the electron is located on the B atom.
 Addition of a second boron has the same effect on the SU center of Si$^2$ \textendash~its lone pair disappears completely.
 The relocation of the second electron results in a further decrease of E$_m$.
 In the As-doped case, the relative arrangement of WCs with SD is similar to the intrinsic case. 
 SU Wannier centers display a more complicated figure: the WC nearby Si$^2$ is suppressed and all three silicon atoms are bond to each other (Fig.~\ref{fig:dopant_sad}~\textit{e-f}). 
 An additional electron provided by a second As atom leads to an almost identical symmetric state for SD (Fig.~\ref{fig:dopant_sad}~\textit{g-h}).
 The corresponding electronic rearrangement results in a slight decrease of the  E$_m$.
 The above Wannier analysis demonstrates that the  E$_m$ reduction is driven by a charge transfer (hole or electron) between the dopant and the oxygen at the saddle configuration.
 
 In addition, the result for B doping are in line with the second hypothesis of Zeng \textit{et al.}\cite{Zeng2011} 
 However, the doping affects more the migration of oxygen monomers than the migration of dimers themselves.
 The formation of dimers thus speeds up by an increase in monomer diffusivity. 
 This has an indirect effect on observed oxygen diffusion by increasing the concentration of dimers while keeping the activation energy constant.
 With this mechanism, the B-enhanced oxygen diffusion originates from an indirect kinetic effect.
 The reported experimental migration energies\cite{Zeng2011, Takeno2000, Murphy2006JAP, Murphy2006MSE}, which are  in the same  1.4-1.5 eV range for both doped and undoped samples, validate the proposed mechanism for B-doping where the oxygen dimers remains the primary migrating species in doped samples. 
 Moreover our DFT calculations clearly show that their migrating properties cannot be significantly affected  by strain, doping type or rate. 
 In other words, we propose that the observed enhancement in p-doped silicon (B, Al and Ga doping) occurs due to the enhanced monomer diffusion, which raises the number of dimers. 
 
 Interestingly, the retarded diffusion in n-doped silicon cannot be attributed to the opposite mechanism (\textit{i.e.} decrease of dimer concentration due to a kinetic effect on the monomer) or to the dimer diffusion itself. 
 However, in the particular case of Sb doping a strong or even complete reduction of O$_{2i}$ concentration can be inferred from obtained negative binding energy (see Table \ref{tab:o_eact}). 
 We also observe that Sb atom traps the oxygen monomer with a trapping energy of 0.52~eV, whereas no trapping was observed for the other dopants for the same dopant-O$_i$ distance (15~\AA).
 Thus we suggest that in the heavily Sb doped case the primary migrating particle is the oxygen monomer.
 Following the analysis on concurrent diffusion \cite{Caliste2006, Pochet2012}, we can infer that the effective activation energy is a function of the migration and binding energies of O$_i$ and O$_{2i}$, as well as the O-dopant trapping or binding energies.
 The exact quantification of this effective diffusion would require an exhaustive study of the sink strength of the dopant, including the possible formation of oxygen-dopant complexes. 
 This is beyond the scope of the present study that aims to identify the involved mechanisms only. 
 Still a lower bond of 2.76 eV, can be given for antimony case as the sum of monomer migration and trapping energies. 
 This value is close to the reported increase of migration energy in Sb-doped samples \cite{Takeno2000}. 
 With this mechanism, the Sb-retarded oxygen diffusion originates from two thermodynamic effects coming together: \textit{i)} negative binding energy of O dimers, and \textit{ii)} long-range trapping of O monomers by Sb atoms.
 Such trapping mechanism will lead to an increased activation energy. 
 Moreover the retardation should be sensitive to the ratio between dopant and oxygen and should vanish at high enough temperature to detrap oxygen from the dopant.

 Up to now, we have proposed two different mechanisms to explain both enhanced and retarded diffusion, which are in line with the reported experimental data for B\cite{Zeng2011} and Sb \cite{Takeno2000}, respectively. 
 Based on our DFT calculations, we have attributed the enhanced oxygen diffusion to reduced monomer migration energy, and the retarded oxygen diffusion to oxygen trapping close to a dopant site. 
 In real samples, the proposed mechanisms can appear at the same time and conclusions derived from the experiments might be contradictory depending on the employed techniques. 
 Indeed, an enhanced oxygen diffusion will accelerate the formation of complexes which further trap oxygen and thus reduce the overall diffusivity. 
 Such an implication could explain the apparently contradictory results in the case of B-doping \cite{Zeng2011} since stable B-O complexes have been identified in the LID literature \cite{Adey2003, Voronkov2010, Schmidt2004}. 
 
 To summarize, we have proposed a consistent picture at the atomistic level of low-temperature diffusion of oxygen monomers and dimers under strain and heavy doping conditions.
 Our DFT calculations show that the oxygen dimer remains the primary migrating species for most doping cases and that its diffusing properties cannot be significantly affected by strain, doping type or rate.
 On the other hand, under high doping rates, we propose two mechanisms (one kinetic and one thermodynamic) that affect the properties of the oxygen monomer, thus creating more or less transient O$_{2i}$ species to diffuse in the crystal. 
 The proposed mechanism for enhanced diffusion is an alternative explanation to the model proposed by Adey \textit{et .al} \cite{Adey2004} for the enhanced kinetics of B-O complexes formation, which are found to be responsible for the light induced degradation \cite{Schmidt2004}.

\begin{acknowledgments}
This work was funded by the French National Research Agency through the BOLID Project ANR-10-HABISOL-001. Calculation time was provided by the French GENCI Agency under Project No. t2011096107.
\end{acknowledgments}

\end{document}